\documentclass{elsart}

\begin{document}
\runauthor{Leszek {\L}ukaszuk}
\begin{frontmatter}
\title{Sum Rules for Parity Violating Compton Amplitudes.}
\author{Leszek {\L}ukaszuk}
\address{Andrzej Soltan Institute for Nuclear Studies, Hoza 69, 00-681
Warsaw,Poland}

\begin{abstract}
Sum rules for parity violating spin polarizabilities are derived
and discussed.
They hold both for hadron and nuclear stable targets of arbitrary spin
and
are exact
in strong interactions. Examples of applications to the cases of proton
and
deuteron
targets are given. The legitimacy of the dispersive approach in the
Standard Model
is discussed.
\end{abstract}
\end{frontmatter}

\newpage \section{Introduction}
The recent revival~\cite{r1,r2,r3,r3a,r4}
 of interest in the weak part of
photon - hadron  interactions
seems to be closely connected with the advent of intense polarized
beams of photons ~\cite{r5,r6,r6a,r6b,r7}. Generally it can be expected
that future experiments will be a good source of information in the
theoretically
difficult domain of low energy hadronic structure. For example the
threshold
production asymmetries in pion photoproduction
on proton ~\cite{r1,r2} would test the consistency of the weak ${NN
\pi}$ coupling ,
$h^{(1)}_{\pi NN}$ with the "best fit" coupling scheme ~\cite{r12}.
Similiar expectations are connected with low energy Compton scattering
~\cite{r3,r3a} on a proton. The $h^{(1)}_{\pi NN}$ problem is acute
because results
from different nuclear ~\cite{r8,r8a} and atomic ~\cite{r9} physics
experiments
seem to indicate quite different values of $h^{(1)}_{\pi NN}$.
The theoretical interpretation of future experiments on more elementary
targets
should
be easier, of course.
At the moment neither the large nor the small $h^{(1)}_{\pi NN}$  option
can be
excluded and it is possible that this constant could
be much smaller ~\cite{r2,r4} then the so called ``best value''
~\cite{r12}
; then the short distance (in comparison with effective $\pi$ exchange)
Parity Violating (p.v.) contributions should be
larger than those from the set of ``best values'' ~\cite{r12} - here a
test
of the importance of such contributions would be the photodisintegration
of
deuterons ~\cite{r4}. Both real and virual photon -initiated effects are
of
interest
and have been considered in the
literature ~\cite{r1,r2,r3,r3a,r4,r16,r16a}. Experimentally these cases
correspond to photon
and electron -initiated collisions, respectively.
A convenient feature of (real) photon initiated
collisions is the absence of direct $Z$ exchange between projectile and
target so it is a unique situation where the p.v. structure of the
electromagnetic current itself is
singled out without further elaborations. On the other hand
disentangling virtual photon p.v. contributions from electroproduction
seems to be difficult - already at $Q^2 > 0.1 GeV^2$  it's contribution to the
measured asymetry is a few percent of the neutral current's contribution
~\cite{r16}. This situation reflects the fact that p.v. interference 
terms involving only electromagnetic currents  must contain an additional 
photon propagator so an extra $(\alpha)$ factor appears  when compared with 
terms  which contain electromagnetic
and neutral current exchanges (Only for $Q^2\rightarrow 0$ will
the additional photon propagator dominate - for $\pi$ meson
electroproduction it takes place below $Q^2$ of the order of $10^{-3}
GeV^2$ ~\cite{r16}). Having this situation in mind we shall limit ourselves to
real photons only in the further discussion so that in the language of
ref.~\cite{r16}  we shall confine ourselves to the p.v. electromagnetic and
nuclear interaction at $Q^2 = 0$. Model dependent estimates
of these interactions lead to asymmetries of the order of $10^{-7}$ for
low energy photoproduction ~\cite{r1,r16} (``best fit `` coupling scheme).
The Compton amplitude asymmetries have been theoretically estimated in
the case of proton target as $10^{-8}$
within the one-loop approximation of heavy-baryon chiral perturbation
theory (HB$\chi$PT) ~\cite{r3,r3a} . As it was shown ~\cite{r21,r21a}  that
p.v. effects at low energies,
$\omega$ , do not contribute to the static terms (i.e. they are at least
of order $O(\omega^2)$) in this process, we can expect that these p.v.
asymmetries and p.v. spin polarizabilities should be closely related.
Therefore it is sensible
to question the credibility of HB$\chi$PT in the context of parity
conserving (p.c.) spin
polarizabilities of the nucleon ~\cite{r24}; in fact the
HB$\chi$PT expansion was shown to be an unstable method for the
calculation of these polarizabilities ~\cite{r22,r221,r222,r223}.
So in the p.v. case, apart from the uncertainty of couplings, this
kind of unstability cannot be ruled out. It would be therefore
appropriate to apply the
dispersive approach, successful in p.c. analysis ~\cite{r23,r5}, using a
low energy multipole expansion and in such a way limiting the number of
adjustable parameters - however, in p.c. analysis it was shown that
apart from general theoretical input, quite acurate experimental data are
needed to determine these parameters ~\cite{r23}. We have no comparable data for
the p.v. analysis at the
moment, so it seems sensible to try to correlate different p.v. hadronic
or nuclear
observables via model independent relations such as dispersive sum
rules, in analogy with those applied to the spin dependent p.c. domain
~\cite{r37a,r37,r24,r241}.
We shall limit ourselves to the sum rules
for forward amplitudes, as in this case relations between physically
measured cross sections are most transparent. In what follows we shall discuss
the legitimacy of the relevant dispersion relations,
apply them to Compton scattering on
protons and deuterons using existing models for p.v. in photoproduction
~\cite{r1,r16} and photo-disintegration ~\cite{r4} .
We shall also discuss the problem of subtractions in the context of the
eventual
use
of superconvergence assumptions.
Tentative predictions made under such assumptions will be formulated.

\section{Asymptotic States in Standard Model} \setcounter{equation}{0}

As we want to discuss dispersion formulae for collision amplitudes, it
is a
suitable place to ask to what  degree the usual properties of these
amplitudes
(existence of asymptotic states and of  interpolating local fields) are
exhibited in the Standard Model. The asymptotic states have to
correspond to a
Fock space of stable particles, so we are left with photons, electrons,
neutrinos
(at least the lightest one ), protons and stable atomic ions. Let us
mention
here
that the existence of unstable fields is a source of
concern in Quantum Field Theory ~\cite{r27,r27a}. Next, each stable
particle
should
correspond to an irreducible Poincare (unitary) representation and here
trouble
appears with
charged particles ~\cite{r28,r28a}. This is connected with QED infrared
radiation
and a well defined procedure exists in perturbative calculus only.
This is the reason why our considerations concerning Compton amplitudes
will
be
limited to the order $\alpha$ in the p.c. part and to order $\alpha^2$
in the p.v.
part (they are infrared safe and at low energies are $\alpha G_F$ order
contributions).
Still we are left with the problem of asymptotic states and
interpolating
fields in the QCD part of the SM - we shall rely on the results of
Oehme:
''the analytic properties of physical amplitudes are the same as those
obtained on the basis of an effective theory involving only the
composite,
physical fields'' ~\cite{r30} (in other words confinement does not spoil
the old
axiomatic proofs for hadronic interactions ~\cite{r31} ).

\section{Parity Violating Compton Sum Rules for Arbitrary Target}
\setcounter{equation}{0}
\subsection{Dispersion Relations and Low Energy Behaviour}
Working in the lowest electroweak order it is reasonable to abandon C, P
invariance only and keep in our considerations T-invariance. The strong 
part of the interactions is taken without any approximation. The analyticity 
of the forward Compton amplitude and crossing properties follow from the 
typical steps sketched below. We start from LSZ - derived 
formula ~\cite{r31a} for the forward Compton amplitude;
for convenience we choose the target in the lab frame and drop
inessential for the further argument contact terms, so that we get ~\cite{r31a}

\begin{equation}
S_{fi} = I + i (2\pi)^4 \delta_{4}(P_{f} - P_{i})M_{fi}
\label{eq1.1}
\end{equation}

\begin{equation}
M_{fi} = e^2 \overline{\epsilon^{\mu}_{f}}\epsilon^{\nu}_{i} T_{\mu\nu}
\label{eq1.2}
\end{equation}

\begin{equation}
T_{\mu\nu}(q,M,s_{f},s_{i}) = i \int d^4 x e^{i\omega (x_0 -
\vec{n}\vec{x})}\Theta(x_0)\langle
M,s_f|[j_{\mu}(x),j_{\nu}(0)]|M,s_i\rangle
\label{eq1.3}
\end{equation}

where $\vec{q}=\omega\vec{n}$ is the photon's momentum and with z-axis
taken in
it's direction,$\vec{n}$, $s_i, s_f$ is the target's initial (final)
z-component
of
spin, $\epsilon_ i, \epsilon_f$ denote initial (final) circular
polarization
states of the photon. Let us work in radiation gauge so only the
$\mu,\nu = 1,2$
components of $j_{\mu},j_{\nu}$ contribute.
In what follows we never use parity conservation.\\
The analyticity in the upper complex half-plane of $\omega$ follows from
the
fact that the retarded commutator in ~(\ref{eq1.3}) vanishes for $x_0
<0$ and ,
due to causality, vanishes also for $x_0 < | z|$. For $Re\omega >0$
approaching the
real axis we get the physical Compton amplitude specified by
~(\ref{eq1.2}).
For $Re\omega <0$ the limiting amplitude can be obtained by applying
complex
conjugation to ~(\ref{eq1.3}) and exploiting invariance of the matrix
elements
with respect to rotations; here rotation around the y or x-axis by angle
$\pi$
should be used. The result, independent of P,C, T invariances, reads

\begin{equation}
M^{s_f,s_i}_{h_f,h_i}(-Re\omega + i\epsilon) =
\overline{M^{s_i,s_f}_{-h_f,-h_i}(Re\omega + i\epsilon)}
\label{eq1.4}
\end{equation}
Demanding T invariance we get

\begin{equation}
M^{s_f,s_i}_{h_f,h_i}(-Re\omega + i\epsilon) =
\overline{M^{s_f,s_i}_{-h_i,-h_f}(Re\omega + i\epsilon)}
\label{eq1.5}
\end{equation}

In what follows we shall be interested in coherent amplitudes only
(i.e. $s_i=s_f$ and $h_i=h_f$), suitable for sum rules,  as their
imaginary
parts are proportional to the total cross sections.
In this case ~(\ref{eq1.4})  and ~(\ref{eq1.5})  are eqiuvalent, so the
demand
of T invariance is not necessary, however, this invariance will be used
below in the estimates
of the low energy behaviour. We shall use abbreviated names $f$
for these amplitudes

\begin{equation}
f_{s,h}(\omega)=M^{s,s}_{h,h}(\omega)
\label{eq1.6}
\end{equation}

so that
\begin{equation}
f_{s,h}(-Re\omega + i\epsilon) = \overline{f_{s,-h}(Re\omega +
i\epsilon)}
\label{eq1.7}
\end{equation}

We shall use amplitudes \cite{r32} normalized such that, for any target

\begin{equation}
Im f_{s,h}(\omega) = \omega \sigma^{T}_{s,h}(\omega)
\label{eq1.7a}
\end{equation}

Then analyticity, crossing ~(\ref{eq1.7}) and unitarity lead, through
Hilbert
formulae to the dispersion relations for these amplitudes

\begin{equation}
Re f_{s,h}(\omega)=\frac{1}{\pi} P \int_{\omega_{th}}^{\infty}\frac{
\omega' \sigma^T_{s,h}(\omega')}{\omega'-\omega}d\omega' +
\frac{1}{\pi}\int_{\omega_{th}}^{\infty}\frac{
\omega' \sigma^T_{s,-h}(\omega')}{\omega'+\omega}d\omega' + (subtr.)
\label{eq1.8}
\end{equation}

On the other hand any amplitude $f$ can be written as
\begin{equation}
f_{s,h}=f^+_{s,h} + f^-_{s,h}
\label{eq1.9}
\end{equation}

where  $f^+$, $f^-$   are p.c. and p.v., respectively,
\begin{equation}
f^{\pm}_{s,h} = \frac{1}{2}(f_{s,h} \pm f_{-s,-h})
\label{eq1.10}
\end{equation}

There exist proofs of low energy QED theorems for targets of any spin
~\cite{r32,r32a} up to $O( \omega)$ terms.
Explicit proof that the p.v. amplitudes of
SM are of order $O( \omega^2)$ has been given for spin $\frac{1}{2}$
~\cite{r21,r21a}. I learned from I.B.Khriplovich that this result holds
for targets of any spin if one neglects, as we do, T-violation;
the reason is that the leading (larger than $O(\omega^2)$) low energy
behaviour
comes from the pole diagrams and that in this case the only p.v.
coupling of a
real photon involves electric dipole moment ~\cite{r33}.
Therefore we can write, for a target of any spin

\begin{equation}
M^{s_f,s_i}_{h_f,h_i}(\omega \rightarrow 0) = \delta_{h_f,h_i}
\delta_{s_f,s_i} f_{s_i, h_i}^{(+)LET} + O(\omega^2) =
M^{(+)}+O(\omega^2)
\label{eq1.11}
\end{equation}

with $f^{(+)LET}$ known from the p.c. Low Energy Theorems
~\cite{r32,r32a}
and

\begin{equation}
M^{(\pm) s_f,s_i}_{h_f,h_i} = \frac{1}{2}(M^{s_f,s_i}_{h_f,h_i}\pm
M^{-s_f,-s_i}_{-h_f,-h_i})
\label{eq1.12}
\end{equation}

so that

\begin{equation}
f^-_{s,h}(\omega)|_{\omega\rightarrow 0}=O( \omega^2)
\label{eq1.13}
\end{equation}

Hence for any target in
the limit $\omega \rightarrow 0$ the ratio

\begin{equation}
A(s_i,h_i) = \frac{ \sum_{s_f,h_f}(|M^{s_f,s_i}_{h_f,h_i}|^2 -
|M^{s_f,-s_i}_{h_f,-h_i}|^2)}{4 f_{s_i, h_i}^{(+) LET}}
= f^{-}_{s_i,h_i} + O(\omega^4)
\label{eq1.14}
\end{equation}

measures the parity violating part of the forward amplitude.
This came out to be simple for any target due to the diagonal form of
$M^+$ at
low
energies
(comp. ~(\ref{eq1.11})). \\
It will be convenient to consider  p.v. amplitudes averaged over the
spin of the target

\begin{equation}
f^{(-)\gamma}_h = \frac{1}{2 S+1} \sum_{s_i}f^{-}_{s_i,h}
\label{eq1.16}
\end{equation}

and averaged over the photon's helicity

\begin{equation}
f^{(-)tg}_s = \frac{1}{2}(f^-_{s,+1} + f^-_{s,-1})
\label{eq1.18}
\end{equation}

These amplitudes are expressed by integrals over relevant
differences of the total cross sections (comp. ~(\ref{eq1.8})).
\begin{equation}
Re f^{(-)\gamma}_h =\frac{\omega}{\pi} P
\int_{\omega_{th}}^{\infty}\frac{
\omega'}{\omega'^2-\omega^2}(\sigma^T_h - \sigma^T_{-h})d\omega' +
(subtr.)
\label{eq1.19}
\end{equation}

where
\begin{equation}
\sigma^T_{h} =\frac{1}{2 S+1} \sum_{s_i}\sigma^T_{s_i,h}
\label{eq1.19a}
\end{equation}

and
\begin{equation}
Re f^{(-)tg}_s =\frac{1}{\pi} P \int_{\omega_{th}}^{\infty}\frac{
\omega'^2}{\omega'^2-\omega^2}(\sigma^T_s - \sigma^T_{-s})d\omega' +
(subtr.)
\label{eq1.20}
\end{equation}

where
\begin{equation}
\sigma^T_s = \frac{1}{2}(\sigma^T_{s,+1} +\sigma^T_{s,-1})
\label{eq1.21}
\end{equation}

\subsection{Sum Rules for p.v. spin polarizabilities}
As the p.v.  amplitudes considered in our approximation are infrared
safe we are entitled to expect that the high energy growth of 
forward amplitudes will be at most $\omega (ln \omega)^2$
as results from general principles for finite range interactions
~\cite{r31}. Condition ~(\ref{eq1.11}) means that for a target of 
any spin  no arbitrary constants appear in the dispersion formulae 
for $f^{(-)tg}$, $f^{(-)\gamma}$
if the subtraction point is taken at $\omega =0$, therefore

\begin{equation}
Re f^{(-)tg}_s =\frac{\omega^2}{\pi} P \int_{\omega_{th}}^{\infty}\frac{
\sigma^T_s - \sigma^T_{-s}}{\omega'^2-\omega^2}d\omega' = - 4 \pi
\omega^2
a_{s}^{(-)tg}(\omega)
\label{eq1.22}
\end{equation}

and

\begin{equation}
Re f^{(-)\gamma}_h = \frac{\omega^3}{\pi}
P \int_{\omega_{th}}^{\infty}\frac{\sigma^T_h - \sigma^T_{-h}}
{\omega'(\omega'^2-\omega^2)}d\omega' = - 4 \pi \omega^3
a_{h}^{(-)\gamma}(\omega)
\label{eq1.23}
\end{equation}

For $\omega \rightarrow 0$ eqns.(\ref{eq1.22}, \ref{eq1.23}) yield sum
rules
for p.v. forward polarizabilities $a_{s}^{(-)tg}(\omega = 0)$,
$a_{h}^{(-)\gamma}(\omega = 0)$ defined in analogy with p.c. forward
spin
polarizabilities ~\cite{r5}

\begin{equation}
a_{s}^{(-)tg}(0) = \frac{1}{4 \pi^2}  \int_{\omega_{th}}^{\infty}\frac{
\sigma^T_{-s} - \sigma^T_{s}}{\omega'^2} d\omega'
\label{eq1.23a}
\end{equation}

\begin{equation}
a_{h}^{(-)\gamma}(0) = \frac{1}{4 \pi^2}
\int_{\omega_{th}}^{\infty}\frac{\sigma^T_{-h} - \sigma^T_{h}}
{\omega'^3}d\omega'
\label{eq1.23b}
\end{equation}

\subsection{Superconvergence Hypothesis}
If we assume superconvergence of
the type $\frac{f(z)}{z} \rightarrow 0$ at infinity
for the asymmetric amplitude ~(\ref{eq1.23}), then
the p.v. analogue of DHG ~\cite{r37a,r37} is obtained

\begin{equation}
\int^{\infty}_{\omega_{th}}\frac{
\sigma^T_h - \sigma^T_{-h}}{\omega'}d\omega' = 0
\label{eq1.28}
\end{equation}

It is natural to question the legitimacy and consequences of such an
assumption for the p.v. amplitude $f^{(-)\gamma}$. 
Consequences of this formula will be mentioned
in the context of model dependent applications in the next section. The
check whether at least in the perturbative QCD regime the relevant
contributions to the total
cross sections asymptotically cancel for a given target in
~(\ref{eq1.28})has not yet been done.
One should calculate a few different processes which
might conspire to give vanishing overall difference of the total cross
sections. It might also happen that the non perturbative regime plays an
equal or essential role. The seemingly plausible conjecture that spin dependent
contributions to the integrated cross sections asymptotically vanish for any reaction,
need not be true. We have checked ~\cite{r34} that a class
of $\gamma$-induced processes with polarized target (proton),
namely those with the production of additional quark antiquark pair via
unpolarized photon structure ~\cite{r35} yields a non vanishing (in fact 
slowly rising)contribution to the difference $\sigma^{T}_{\frac{1}{2}} -
\sigma^{T}_{-\frac{1}{2}}$ if present-day parton distributions are used
~\cite{r36}. Despite this warning we shall study the consequences of
~(\ref{eq1.28}) in the next section.\\

\section{Examples of Applications} \setcounter{equation}{0}
\subsection{Proton Target}
The  p.v. Compton amplitude can be written in the c.m.s. as
~\cite{r3,r3a}
\begin{eqnarray}
M^{(-)s_f,s_i}_{h_f,h_i}(\vec{k},\vec{k'})&=&
\overline{N_{s_f}} [F_1\vec{\sigma}\cdot(\hat{\vec{k}} +\hat{\vec{k'}})
\vec{\epsilon_i}\cdot\overline{\vec{\epsilon'_f}} -
F_2 (\vec{\sigma}\cdot\overline{\vec{\epsilon'_f}}
\hat{\vec{k'}}\cdot\vec{\epsilon_i} +
\vec{\sigma}\cdot\vec{\epsilon_i}
\hat{\vec{k'}}\cdot\overline{\vec{\epsilon'_f}}) \nonumber\\
 &-&F_3 \hat{\vec{k}}\cdot\overline{\vec{\epsilon'_f}}
 \hat{\vec{k'}}\cdot\vec{\epsilon_i}\vec{\sigma}\cdot(\hat{\vec{k}} +
\hat{\vec{k'}}) - i
F_4\vec{\epsilon_i}\times\overline{\vec{\epsilon'_f}}
\cdot(\hat{\vec{k}} +\hat{\vec{k'}})] N_{s_i}
\label{eq1.29}
\end{eqnarray}

so that
\begin{equation}
f^{(-)p}_{\frac{1}{2}} = 2 F_1 =O(\omega^2)
\label{eq1.30}
\end{equation}
\begin{equation}
f^{(-)\gamma}_{h=+1} = -2 F_4 =O(\omega^3)
\label{eq1.31}
\end{equation}

The HB$\chi$PT analysis ~\cite{r3,r3a} provides values of coefficients
$F_1,F_4$
\begin{equation}
F_1|_{\omega\rightarrow 0} = -\frac{e^2}{M}(\frac{\omega}{m_{\pi}})^2
\frac{M}{F_{\pi}}\frac{g_{A}h^{(1)}_{\pi NN}}{24\sqrt{2}\pi^2}
\label{eq1.32}
\end{equation}
\begin{equation}
F_4|_{\omega\rightarrow 0} = \frac{e^2}{M}(\frac{\omega}{m_{\pi}})^3
\frac{m_{\pi}}{F_{\pi}}\frac{g_{A}h^{(1)}_{\pi
NN}\mu_n}{24\sqrt{2}\pi^2}
\label{eq1.33}
\end{equation}
where $F_{\pi} = 93 $MeV, ~$g_A= 1.26$, ~$\mu_n = -1.91$, ~$h^{(1)}_{\pi
NN}
\simeq 5\cdot 10^{-7}$ in the ``best fit'' parametrization, $M$ is the
nucleon mass. On the other hand there are theoretical results for p.v. effects in
near threshold photoproduction ~\cite{r1,r2,r16}.
These values could be used in our relations ~(\ref{eq1.22}, \ref{eq1.23}). 
Denoting
\begin{equation}
Re f_{s=\frac{1}{2}}^{(-)p} = - \frac{e^2}{M}
\beta^- (\frac{\omega}{m_{\pi}})^2|_{\omega\rightarrow 0}
\label{eq1.34}
\end{equation}
\begin{equation}
Re f_{h=+1}^{(-)\gamma} = - \frac{e^2}{M}
\gamma^- (\frac{\omega}{m_{\pi}})^3|_{\omega\rightarrow 0}
\label{eq1.35}
\end{equation}
we have
\begin{equation}
a_{\frac{1}{2}}^{(-)tg}(0) = \frac{\alpha}{M m_{\pi}^{2}} \beta^{-} =
3.2 \beta^{(-)} 10^{-3} [fm]^3
\label{eq1.35a}
\end{equation}

\begin{equation}
a_{1}^{(-) \gamma}(0) = \frac{\alpha}{M m_{\pi}^{3}} \gamma^{-} =
4.5 \gamma^{(-)} 10^{-3} [fm]^4
\label{eq1.35b}
\end{equation}

with

\begin{equation}
\beta^- =-(\frac{e^2}{M m^2_{\pi}})^{-1}
\frac{1}{\pi}\int_{\omega_{th}}^{\infty}\frac{
\sigma^T_{s=\frac{1}{2}}-\sigma^T_{s=-\frac{1}{2}}}
{\omega'^2}d\omega'
\label{eq1.36}
\end{equation}
\begin{equation}
\gamma^- =-(\frac{e^2}{M m^3_{\pi}})^{-1}
\frac{1}{\pi}\int_{\omega_{th}}^{\infty}\frac{
\sigma^T_{h=+1}-\sigma^T_{h=-1}}
{\omega'^3}d\omega'
\label{eq1.37}
\end{equation}

while $\beta_{\chi}, \gamma_{\chi}$ coming from
~(\ref{eq1.32},\ref{eq1.33})
 are

\begin{equation}
\beta^-_{\chi} = \frac{M}{F_{\pi}} \frac{g_A h^{(1)}_{\pi
NN}}{12\sqrt{2}\pi^2}
\simeq 4 \cdot 10^{-8}
\label{eq1.38}
\end{equation}
\begin{equation}
\gamma^-_{\chi} = \mu_n \frac{m_{\pi}}{F_{\pi}} \frac{g_A h^{(1)}_{\pi
NN}}
{12\sqrt{2}\pi^2}
\simeq -1 \cdot 10^{-8}
\label{eq1.39}
\end{equation}

 We shall use cross sections from ~\cite{r1,r16} in our sum rules,
compare
consistency and finally, assuming superconvergence ~(\ref{eq1.28})
discuss
posible consequences. Let us start with the HB$\chi$PT approach to
photoproduction;
taking dominant at threshold terms in the effective lagrangian of
reference
~\cite{r1}
we integrate them up to $\omega= 200$ MeV i. e. in the region where
HB$\chi$PT
should be reliable. The results are $1 \cdot 10^{-8}$
for $\beta^-$   and $-4 \cdot 10^{-9}$ for $\gamma^-$ ,
to compare with $4 \cdot 10^{-8}$ and $-1 \cdot 10^{-8}$
for $\beta^-_{\chi}$ and $\gamma^-_{\chi}$, respectively.
This means that in ($\gamma p \longrightarrow \gamma p$) HB$\chi$PT
calculations
\cite{r3,r3a} contributions from much higher energies have been
involved.
Indeed, extrapolating the threshold behaviour up to $1$ GeV and
inserting into
~(\ref{eq1.36}), ~(\ref{eq1.37}) we get values
which compare well with results from ~\cite{r3,r3a}.
However, there is no reason to assume the validity of threshold type
behaviour
in such a large region. Therefore we turn to an analysis ~\cite{r16}
where
elaborated Born type exchanges (with resonances and form factors
considered)
were put
together. We shall consider the ``best fit'' predictions contained in
figs 11-15
of ~\cite{r16}. Using eqs. ~(\ref{eq1.36}),~(\ref{eq1.37}) we get
$-5 \cdot 10^{-9}$ for $\beta^-$ and $-1 \cdot 10^{-8}$
for $\gamma^-$. This is an example of the usefulness of measurements,
not only of
threshold p.v. photoproduction, but low energy Compton asymmetries, too,
in
future experiments, as  they can shed some light on the existence of
structure at higher energies. In the comparison made above the $\beta^-$
obtained differ by an order of magnitude. This is a reflection of the quite
different behaviour of cross sections.
Let us pass now to the superconvergence hypothesis ~(\ref{eq1.28}).
The contributions in the region below $0.55$ GeV we calculate from
~\cite{r16} and obtain the relation

\begin{equation}
\int_{0.55 GeV}^{\infty}\frac{\sigma^T_1 -
\sigma^T_{-1}}{\omega'}d\omega'
\simeq -30 pb
\label{eq1.40}
\end{equation}

If we further assume - by analogy with the gross features of DHG sum
rule saturation - that the necessary contribution comes from the region below
$1$ GeV we get for the average asymmetry in the region ($0.55-1$ GeV) a
value ($-50$ pb).
This might indicate that it is desirable to look for p.v. effects in
this region.

\subsection{Deuteron Target}
We shall consider the model of Khriplovich and Korkin ~\cite{r4} for
p.v. effects in photodisintegration. As emphasized in ref. ~\cite{r4}, this
process may provide a
test for the importance of short distance (in comparison with $\pi$
exchange) Parity Violating
contributions. Here the difference of cross sections,
integrated up to $10$ MeV, yields a rather small contribution  in
natural units (i.e. ~$\frac{\omega}{2.23MeV}$) but compared
with that for proton i.e. in units
$\frac{\omega}{m_{\pi}}$ it is large:

\begin{equation}
Re f_{h=+1}^{(-)\gamma} = - \frac{e^2}{M_D}
\gamma^{-}_{D} (\frac{\omega}{m_{\pi}})^3|_{\omega\rightarrow 0}
\label{eq1.40a}
\end{equation}

with

\begin{equation}
\gamma^{-}_{D} \simeq - 2 \cdot 10^{-4}
\label{eq1.41}
\end{equation}

If we use a dispersion relation close to the disintegration
threshold $\omega \simeq 2.23MeV$, we get

\begin{equation}
- Re f^{(-)\gamma}_{+1}(\omega) \cdot (\frac{e^{2}}{M_D})^{-1}
|_{\omega \rightarrow 2.23MeV} \simeq -1 \cdot 10^{-8}
\label{eq1.42}
\end{equation}

while extrapolation of (\ref{eq1.40a}) would give $-7 \cdot 10^{-10}$.
This
indicates that p.v. effects are strengthened by an order of magnitude
due to the cusp effect.

\section{Concluding remarks}\setcounter{equation}{0}
New sum rules derived in Section3 were found helpful in checking the
consistency of various theoretical approaches (see discussion 
of the example in Section4.1).
Moreover, under the superconvergence hypothesis (see Section.3.3)
p.v. effects in photoproduction far
from the threshold (see Section.4.1) can be estimated from the low
energy data (models).
The subject of superconvergence seems to be a challenge for further
theoretical studies. \\
As our sum rules and dispersive formulae hold for stable nuclear targets
of arbitrary spin, future applications to complex nuclei are of interest. 
In this context the cusp enhancement of p.v. effects on the deuteron (see Section4.2)
indicates that Compton scattering at energies close to the nuclear
inelastic thresholds should be of importance.

\begin{center}
{\bf Acknowledgements}
\end{center}
I would like to thank Professors: Mamoru Fujiwara, Elliot Leader and 
Ziemowid Sujkowski for encouragement and discussions on photon induced 
processes, Youlik Khriplovich for enlightment on LET in SM and 
Jacques Bros for helpful remarks on analyticity problem in QED.
Many thanks to dr Remco Zegers for  helpful correspondance on
asymmetries.

\end{document}